\documentclass[journal,10pt,onecolumn]{IEEEtran}
\usepackage{subfigure}
\usepackage{cite}
\usepackage{amsmath,amssymb,amsfonts}
\usepackage{algorithmic}
\usepackage{graphicx}
\usepackage{textcomp}
\usepackage{float}
\usepackage{subfigure}

\def\BibTeX{{\rm B\kern-.05em{\sc i\kern-.025em b}\kern-.08em
    T\kern-.1667em\lower.7ex\hbox{E}\kern-.125emX}}
\begin{document}
\title{Directional Adaptive MUSIC-like Algorithm under\\ Symmetric $\alpha-$Stable Distributed Noise}
\author{Narong Borijindargoon and Boon Poh Ng
\thanks{The authors are with the School of Electrical and Electronic Engineering, Nanyang Technological University, Singapore (e-mail: narong001@e.ntu.edu.sg; ebpng@ntu.edu.sg)}
}

\maketitle
\thispagestyle{empty}

\begin{abstract}
An algorithm called MUSIC-like algorithm was originally proposed as an alternative method to the MUltiple SIgnal Classification (MUSIC) algorithm for direction-of-arrival (DOA) estimation. Without requiring explicit model order estimation, it was shown to have robust performance particularly in low signal-to-noise ratio (SNR) scenarios. In this letter, the working principle of a relaxation parameter $\beta$, a parameter which was introduced into the formulation of the MUSIC-like algorithm, is provided based on geometrical interpretation. To illustrate its robustness, the algorithm will be examined under symmetric $\alpha-$stable distributed noise environment. An adaptive framework is then developed and proposed in this letter to further optimize the algorithm. The proposed adaptive framework is compared with the original MUSIC-like, MUSIC, FLOM-MUSIC, and SSCM-MUSIC algorithms. A notable improvement in terms of targets resolvability of the proposed method is observed under different impulse noise scenarios as well as different SNR levels. 

\end{abstract}


\section{Introduction}
Symmetric $\alpha-$stable ($S\alpha S$) distribution is a sub class of stable distributions that is practical for modelling noise due to its ability to encompass the behavior of both Gaussian and Non-Gaussian noises (specifically impulse noise), which can be found in real applications such as the atmospheric noise due to thunderstorms, under-ice and shallow water noise in sonar and submarine communication, sea clutter ambient noise, faulty sensors, and other man-made noises \cite{alphastable}. The $\alpha-$stable distribution can be characterized by several parameters $S(\alpha,\beta,\gamma,\mu)$ which are the characteristic exponent $0<\alpha\leq 2$, the skewness parameter $\beta$, the scaling parameter $\gamma$, and the location parameter $\mu$. In this letter, the distribution is assumed to have no skewness ($\beta=0$), and hence the characteristic function of the $S\alpha S$ distribution can be expressed by
$\varphi(t) = \text{exp}(j\mu t-\gamma |t|^{\alpha})$.
The distribution is also assumed to have zero mean ($\mu = 0$). The isotropic $S\alpha S$ distribution is now characterized mainly by $\alpha$ and $\gamma$, where  $\alpha$ indicates the likelihood of outlier occurrence, and $\gamma$ functions in a similar fashion as the standard deviation ($\sigma$) in a Gaussian process. 

The MUSIC algorithm \cite{RO} is a well-known eigenstructure-based method with super resolution performance \cite{MUSICBest} which has been employed in various contemporary applications from automotive radars \cite{radar}, unmanned aerial vehicle (UAV) localization \cite{UAV}, and underwater sonar to problems in joint sparse signal recovery \cite{compressMUSIC,rankaware,subspace} and anomaly detection \cite{DOT,Transcan,EIT}. In certain scenarios where an impulse noise is present, the characteristic exponent is less than 2 ($\alpha<2$), and hence it is known that only moments of order less than or equal to $\alpha$ is finite \cite{alphastable}. Under such circumtances, the model order estimators which are based on the second-order moment such as the Akaike Information Criterion (AIC) and the Minimum Description Length (MDL) are likely to produce inaccurate estimation, and hence performance degradation of the MUSIC algorithm is expected. Several non-iterative methods which can be regarded as 1-step M-estimator (a class of weighted covariance matrix) \cite{Mestimator}  were proposed in the literature to extract valuable information from the data matrix in the case where a complete knowledge of accurate second-order moment is unavailable \cite{FLOM,ROC,SSCM}. The fractional lower order moment (FLOM) technique was investigated and reported to have comparable performance to the robust covariation technique. With $M-$element sensors, each element of an $M\times M$ FLOM matrix can be obtained by $C_{ik} = E[x_{i}(t)|x_{k}(t)|^{p-2}x_{k}^{*}(t)],$
where $1<p<\alpha\leq 2$, $x_{i}(t)$ and $x_{k}(t)$ are the data sample obtained from $i^{th}$ and $k^{th}$ sensors, and  $E[\cdot]$ denotes the expectation operation. The spatial sign covariance matrix (SSCM), an intuitive and effective method, can also be used where each element of an $M\times M$ SSCM can be obtained by $C_{ik} = E[(\sum_{i=1}^{M}|x_{i}(t)|^{2})^{-1/2}x_{i}(t)(\sum_{k=1}^{M}|x_{k}(t)|^{2})^{-1/2}x_{k}^{*}(t)]$.

To circumvent the subspace partitioning requirement altogether, an algorithm called MUSIC-like algorithm can be used as an alternative method \cite{Zhangying}. The additional relaxation parameter ($\beta$) introduced into the algorithm has enabled the algorithm to achieve high resolution performance comparable to the MUSIC algorithm without requiring explicit model order estimation. Theoretical aspect of the algorithm was analyzed in \cite{Vinod}, and experimental studies with real data under controlled environment were conducted in \cite{Sonar,EIT}. In order to generalize the applicability of the MUSIC-like algorithm to a broad range of scenarios, the algorithm will be examined under $S\alpha S$ distributed noise in this letter. A framework for directional adaptive MUSIC-like is then developed and proposed with an objective to optimize its performance. 

Major contributions of this letter can be concisely summarized into three points as follows. Firstly, a geometrical interpretation of the MUSIC-like algorithm is provided as a complimentary to its theoretical counterpart which was provided in \cite{Vinod}. To demonstrate its robustness under different noise distributions (Gaussian and heavy-talied), the algorithm will be examined under $S\alpha S$ distributed noise. Secondly, we show that the algorithm can be further optimized through an adaptive framework where we propose directional adaptive $\beta-$selection method. The difference to its original formulation is reflected in the relaxation parameter $\beta$, which is now direction-dependent where its value can be automatically readjusted corresponds to each look direction instead of a fixed value in the original formulation. Lastly, performance of the MUSIC-like algorithms (fixed $\beta$ and adaptive $\beta$) are compared with the MUSIC algorithm with and without pre-conditioned covariance matrix (weighted covariance matrix).

\section{Problem Formulation}
Consider an \textit{M}-sensor uniform linear array (ULA) with half-wavelength spacing situated in the far field with \textit{K} narrowband signal sources $\mathbf{s}(n)\in C^{K\times 1}$ impinge along the direction $\Theta_{K} = [\theta_{1},\dots,\theta_{K}]^{T}$. The sensor's snapshot $\mathbf{x}(n)$ can be modelled as 
\begin{equation}
\mathbf{x}(n) = \mathbf{A}(\Theta_K)\mathbf{s}(n)+\mathbf{v}(n),
\end{equation}
where $\mathbf{A}(\Theta_K)=[\mathbf{a}(\theta_{1}),\dots,\mathbf{a}(\theta_{K})]\in \mathbb{C}^{M\times K}$ is an array manifold comprises $K$ steering vectors correspond to each source direction, and $\mathbf{v}(n)\in \mathbb{C}^{M\times 1}$ denotes additive uncorrelated noise vector with zero mean. Each steering vector $\mathbf{a}(\theta)$ is a function of direction $\theta$, which can be expressed as 
\begin{equation}
\mathbf{a}(\theta)=\frac{1}{\sqrt{M}}[\text{exp}(j \mathbf{k}_{\theta}^{T}\mathbf{r}_{1}),\dots,\text{exp}(j \mathbf{k}_{\theta}^{T}\mathbf{r}_{M})]^{T},
\end{equation}
where $\mathbf{r}_{i}$ incorporates the location information of the $i^{th}$ sensor, $\mathbf{k}_{\theta}=2\pi f/v\mathbf{u}_{\theta}$ denotes the wave number, and $\mathbf{u}_{\theta}$ is the unit vector along the wave propagation direction. The signal frequency and the propagation speed are denoted as $f$ and $v$, respectively.  Under the assumption that the signal sources are uncorrelated and infinite number of snapshots can be obtained, the covariance matrix $\mathbf{R}=E[\mathbf{X}\mathbf{X}^{H}]$, can be decomposed into two orthogonal subspaces by eigendecomposition. The covariance matrix can be expressed as
$\mathbf{R} = \mathbf{U}_{s}\Lambda_{s}\mathbf{U}_{s}^{H} + \mathbf{U}_{n}\Lambda_{n}\mathbf{U}_{n}^{H}$,
where $\mathbf{U}_{s}\in \mathbb{C}^{M\times K}$ denotes the signal subspace matrix which comprises the eigenvectors correspond to the dominant eigenvalues in the matrix $\Lambda_{s}=\text{diag}\{\sigma_{s_1}^{2}+\sigma_{v_1}^{2},\dots,\sigma_{s_K}^{2}+\sigma_{v_K}^{2}\}$, and $\mathbf{U}_{n}\in \mathbb{C}^{M\times M-K}$ denotes the noise subspace matrix which comprises the eigenvectors correspond to the noise eigenvalues in the matrix $\Lambda_{n}=\text{diag}\{\sigma_{v_{K+1}}^{2},\dots,\sigma_{v_M}^{2}\}$. The pseudospectrum of the MUSIC algorithm can be obtained by $P_{M}(\theta) = 10\log_{10}(1/\mathbf{a}(\theta)^{H}\mathbf{U}_{n}\mathbf{U}_{n}^{H}\mathbf{a}(\theta))$. The MUSIC-like algorithm was proposed in \cite{Zhangying} as an optimization problem for each look direction defined as
\begin{align}\label{MUSIClikealgorithm}
\begin{array}{r l}
    min_{\mathbf{w}} & \quad\quad \mathbf{w}^{H}\mathbf{Rw}\\
    s.t.             & \quad\quad \mathbf{w}^{H}\mathbf{a}(\theta)\mathbf{a}(\theta)^H\mathbf{w}+\beta||\mathbf{w}||^2_2=c ,
\end{array}
\end{align}
where $\mathbf{w}$ is the weight vector solution of the optimization problem in (\ref{MUSIClikealgorithm}), a scalar value $\beta$ is a relaxation parameter, and $c$ is any constant value. It was shown in \cite{Zhangying} that the weight vector solution $\mathbf{w}$ is the eigenvector corresponds to the minimum eigenvalue $\lambda_{min}$ of the generalized eigenvalue problem
$\mathbf{R} \mathbf{w}= \lambda(\mathbf{a}(\theta)\mathbf{a}(\theta)^H+\beta \mathbf{I})\mathbf{w},$
and hence the spatial spectrum of the MUSIC-like algorithm can be obtained by $P_{Mlike}(\theta) = 10\log_{10}(1/|\mathbf{w}^H\mathbf{a}(\theta)|^2).$ The bound for $\beta$ was proposed in \cite{Vinod} as
\begin{equation} \label{bound}
\underbrace{\max_{\theta \in \Theta}\frac{\lambda_{\mathbf{R},min}}{(\mathbf{a}(\theta)^{H}\mathbf{R}^{-1}\mathbf{a}(\theta))^{-1}}}_\text{$\beta_{min}$}<\beta<\underbrace{\min_{\theta \notin \Theta}\frac{\lambda_{\mathbf{R},min}}{(\mathbf{a}(\theta)^{H}\mathbf{R}^{-1}\mathbf{a}(\theta))^{-1}}}_\text{$\beta_{max}$},
\end{equation}
with the choice of $\beta$ to be a value between $\beta_{min}$ and $\beta_{max}$ defined as
\begin{equation}\label{betaold}
    \beta = (1-\xi)\beta_{min}+\xi\beta_{max},
\end{equation}
where $0<\xi<1$ can be chosen by $\xi = \beta_{min}/\beta_{max}$.

\section{Geometrical Interpretation of the Algorithm}
In this section, a geometrical interpretation of the MUSIC-like algorithm is provided. For ease of visualization, the two dimensional plot is used for illustration purposes. In the case of higher dimensions, a line can be extended to a hyperplane, and an ellipse can be extended to a hyperellipsoid. 

\begin{figure}[!ht]
    \centering
    \subfigure[Capon]
    {
        \includegraphics[width=1.0in]{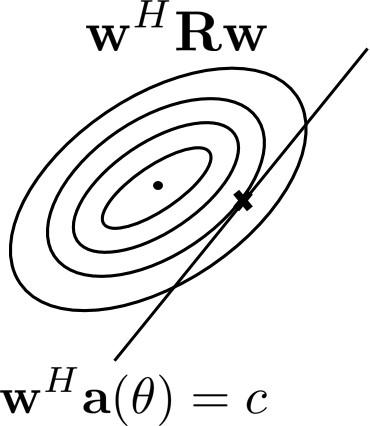}
        \label{CaponContour}
    }
    \subfigure[MUSIC-like]
    {
        \includegraphics[width=1.6in]{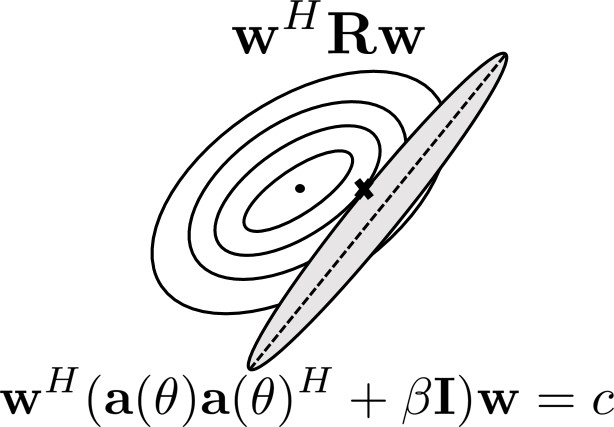}
        \label{MUSIClikeContour}
    }
    \caption{Optimization problems of Capon's beamformer and MUSIC-like algorithm}
    \label{Contours}
\end{figure}

The optimization problem of Capon's beamformer attempts to minimize the output power as $min_{\mathbf{w}}\{\mathbf{w}^H\mathbf{Rw}\}$ while the weight vector solution is required to satisfy the constraint $\mathbf{w}^{H}\mathbf{a}(\theta)=c$ (where $c=1$ is commonly used). The constraint of the Capon's beamformer provides a feasible region where the weight vector solution should belong to, which is on a line (or a hyperplane in high dimensional spaces) as shown in Fig \ref{CaponContour}. In contrast, it can be seen in Fig. \ref{MUSIClikeContour} that the constraint of the MUSIC-like algorithm provides a wider feasible region (ellipsoidal surface) for the weight vector solution to be resided in. This relaxation is in fact promoting the weight vector solution to be resided in the noise subspace, which is a crucial step to obtaining the super resolution performance.

\begin{figure}[H]
    \centering
    \subfigure[Low SNR]
    {
        \includegraphics[width=1.3in]{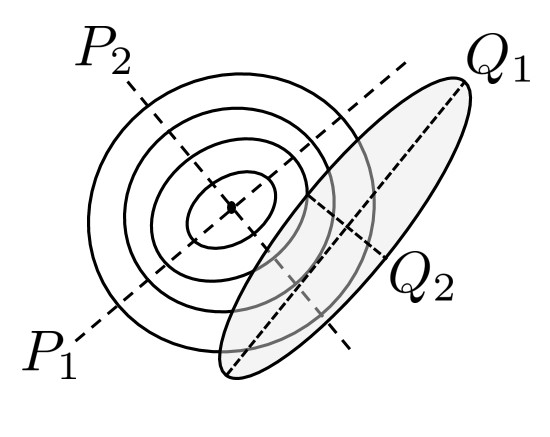}
        \label{MUSIClikeContourFat}
    }\hspace{4mm}
    \subfigure[High SNR]
    {
        \includegraphics[width=1.3in]{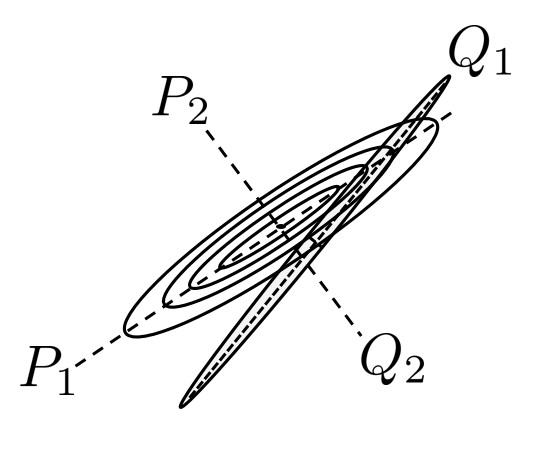}
        \label{MUSIClikeContourSlim}
    }
    \caption{Adaptability of the MUSIC-like algorithm under different SNR scenarios.}
    \label{MUSIClikeContour2}
\end{figure}

\begin{figure}[H]
    \centering
        \includegraphics[width=80mm]{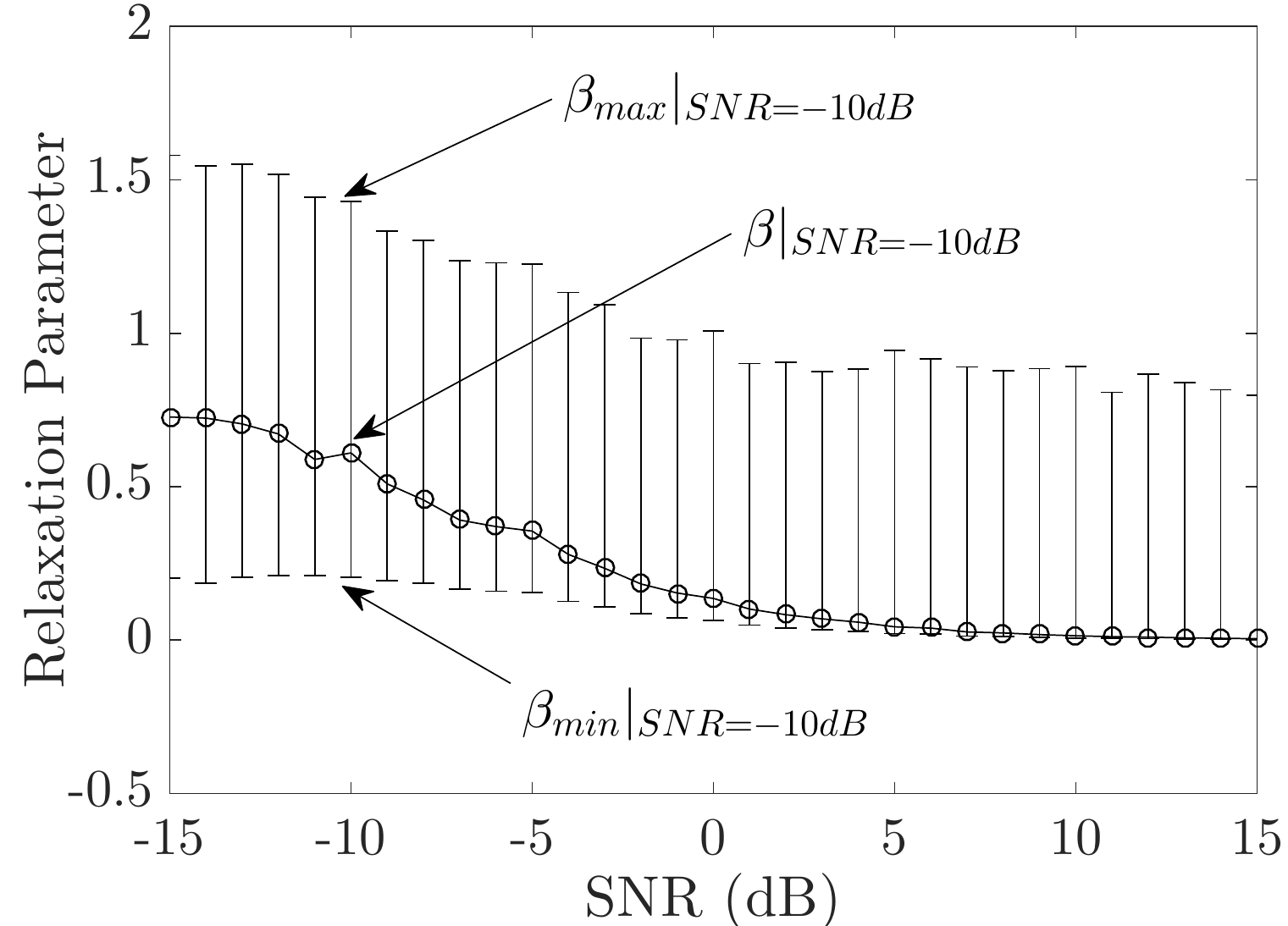}
    \caption{Variation of relaxation parameter under different SNR levels.}
    \label{BetaVariation}
\end{figure}

\noindent From Fig. \ref{MUSIClikeContour}, it can be seen that the $\beta$ parameter plays a key role in regulating the feasible region and the ellipsoidal surface for weight vector solution of the MUSIC-like algorithm. Further details regarding the adaptability of $\beta$ parameter corresponds to the ellipsoidal shape of the covariance matrix is shown in Figs. \ref{MUSIClikeContour2} and \ref{BetaVariation}.


\noindent Note that Fig. \ref{BetaVariation} was obtained from a ULA of $M = 10$ sensors. Two targets were situated at $\Theta_{K} = [50^{\circ}, 110^{\circ}]$, and 200 snapshots of data were obtained. For the case of two dimensional spaces as shown in Fig. \ref{MUSIClikeContour2}, the principle axis $P_{1}$ and the minor axis $P_{2}$ are the first and the second eigenvectors that were extracted from the covariance matrix $\mathbf{R}$ of size $2\times2$. The principle axis $Q_{1}$ and the minor axis $Q_{2}$ are the first and the second eigenvectors that were extracted from the matrix $\mathbf{a}(\theta)\mathbf{a}(\theta)^{H}+\beta\mathbf{I}$ of size $2\times2$. When the SNR is low ($\lambda_{\mathbf{R},1}\approx\lambda_{\mathbf{R},2}$) as shown in Fig. \ref{MUSIClikeContourFat}, both axes ($P_{1}$) and ($P_{2}$) are comparable, and hence by following (\ref{bound}) and (\ref{betaold}), the $\beta$ value is set to a relatively larger value than $\beta_{min}$ as shown in Fig. \ref{BetaVariation}. This large $\beta$ is inflating the ellipsoidal surface accordingly as shown in Fig. \ref{MUSIClikeContourFat}. On the other hand when the SNR is high ($\lambda_{\mathbf{R},1}\gg\lambda_{\mathbf{R},2}$) as shown in Fig. \ref{MUSIClikeContourSlim}, the principle axis ($P_{1}$) is no longer equivalent to the minor axis ($P_{2}$). It can be seen in Fig. \ref{MUSIClikeContourSlim} that the contour plot is now skewed, and by following (\ref{bound}) and (\ref{betaold}), the value of $\beta$ parameter is now approaching the value of $\beta_{min}$ as shown in Fig. \ref{BetaVariation}. This small $\beta$ is deflating the feasible ellipsoidal surface as shown in Fig. \ref{MUSIClikeContourSlim}.

Until recently \cite{Sonar, EIT}, the value of $\beta$ has been set according to the SNR of the obtained data (data-dependent), where it is fixed throughout all look directions ($\beta$ in Fig. \ref{adaptivebeta}). In the next section, $\beta$ will be adaptively readjusted not only according to the SNR (data-dependent) but also throughout each look direction (direction-dependent). This adaptive beta is denoted as $\beta_{\theta}$.

\section{The Working Principle of $\beta_{\theta}$ parameter}
\begin{figure}[H]
    \centering
        \includegraphics[width=80mm]{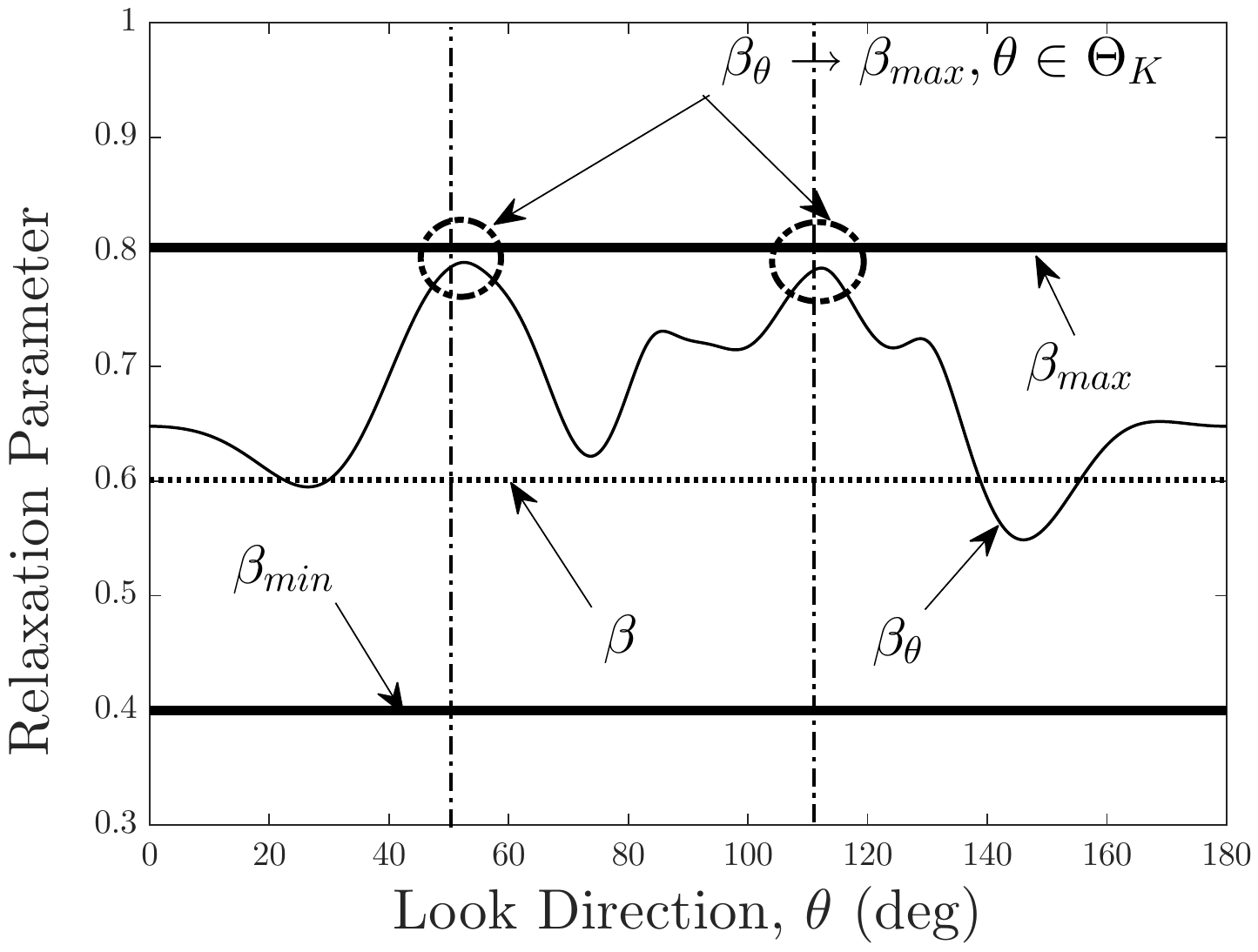}
    \caption{Variation of relaxation parameters correspond to each look direction.}
    \label{adaptivebeta}
\end{figure}

The constraint of MUSIC-like algorithm in (\ref{MUSIClikealgorithm}) can be regarded as an inflated hyperellipsoidal surface where the principle axis of the hyperellipsoid is spanned by a rank 1 matrix $\mathbf{a}(\theta)\mathbf{a}(\theta)^{H}$, and the minor axes are inflated by the $\beta$ value. 
\begin{figure}[!ht]
\centering
\includegraphics[width=3.6in]{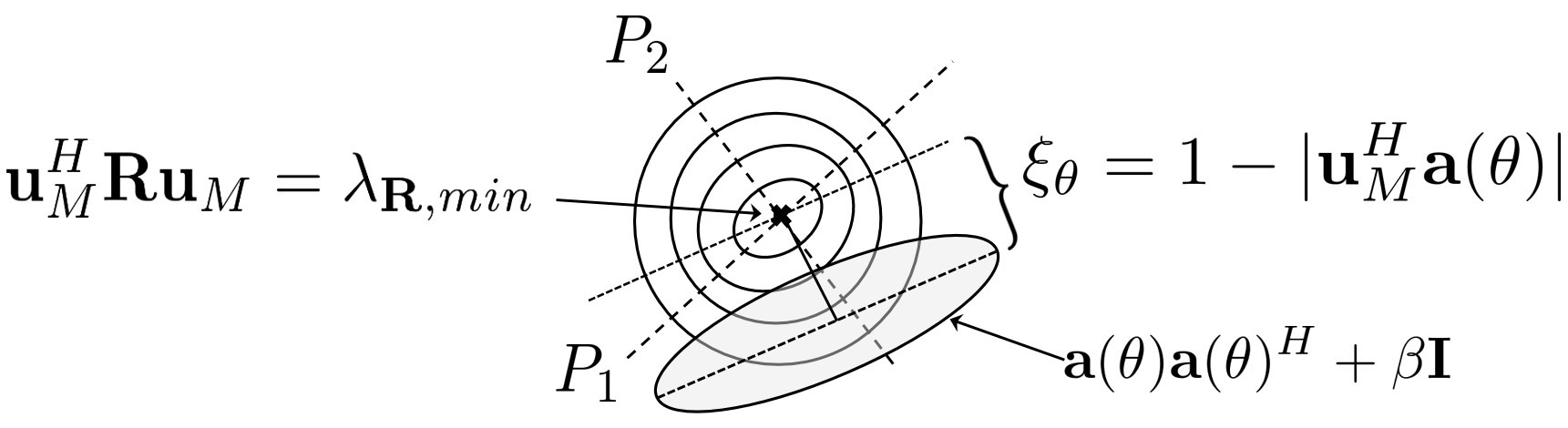}
\caption{The minimum eigenvector corresponds to the minimum eigenvalue of the covariance matrix is used as an anchor point for the distance parameter.}
\label{mineig}
\end{figure}
For each look direction, the distance between the noise subspace and the hyperplane as well as the orientation of the hyperplan itself are varied relative to each steering vector $\mathbf{a}(\theta)$. Since the noise subspace $\mathbf{U}_{n}$ is not known a priori, the eigenvector $\mathbf{u}_{M}$ corresponds to the minimum eigenvalue of the covariance matrix $\lambda_{\mathbf{R},min}$ is used as an anchor point as shown in Fig. \ref{mineig}. 
\begin{figure}[!ht]
    \centering
    \subfigure[]
    {
        \includegraphics[width=1.35in]{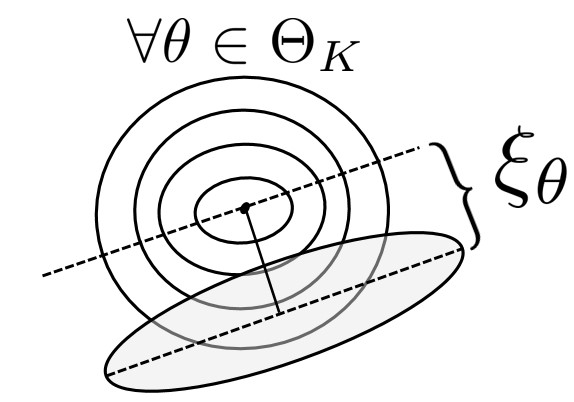}
        \label{alpha1}
    }\hspace{6mm}
    \subfigure[]
    {
        \includegraphics[width=1.2in]{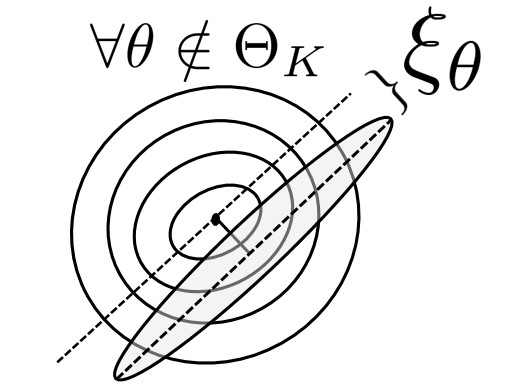}
        \label{alpha2}
    }
    \caption{The distance parameter $\xi_{\theta}$ corresponds to each look direction.}
    \label{alphapic}
\end{figure}
We now propose a distance parameter $\xi_{\theta}$ which approximates the distance between each steering vector and the anchor point. The distance parameter is defined as
\begin{equation}\label{alpha}
\xi_{\theta} = 1-|\mathbf{u}_{M}^{H}\mathbf{a}(\theta)|,
\end{equation}
where $||\mathbf{a}(\theta)||_{2}^{2}=1$, and the value of $\xi_{\theta}$ is varied within the range of $0\leq\xi_{\theta}\leq1$. The new adaptive $\beta_{\theta}$ can be obtained by substituting (\ref{alpha}) into (\ref{betaold}) which can be re-expressed for each look direction as
\begin{equation}
    \beta_{\theta} = \beta_{max}-\delta\beta|\mathbf{u}_{M}^{H}\mathbf{a}(\theta)|,
\end{equation}
where $\delta\beta = \beta_{max}-\beta_{min}$. The working principle of $\beta_{\theta}$ can be summarized as follows. Consider a ULA with $M = 10$ sensors. Two targets are situated at $\Theta_{K}=[50^{\circ},110^{\circ}]$ where 200 snapshots of data were obtained with SNR = -5dB ($\alpha = 2$).  When $\theta\in \Theta_{K}$, $\xi_{\theta}$ admits a relatively large value since $|\mathbf{u}_{M}^{H}\mathbf{a}(\theta)|$ is small. Hence, the corresponding $\beta_{\theta}$ is assigned to a large value as shown in Fig. \ref{adaptivebeta}. With large $\beta_{\theta}$, the ellipsoidal surface is then inflated. In contrast, when $\theta\notin \Theta_{K}$, $\xi_{\theta}$ admits a relatively small value since $|\mathbf{u}_{M}^{H}\mathbf{a}(\theta)|$ is large. The corresponding $\beta_{\theta}$ is now assigned to a relatively small value. With small $\beta_{\theta}$, the ellipsoidal surface is now deflated accordingly.  
The variation of $\xi_{\theta}$ allows $\beta_{\theta}$ to be adaptively readjusted itself for each look direction, and hence the hyperellipsoid is automatically inflated or deflated corresponding to each look direction. On the other hand, the old $\beta$ (marked as $\beta$ in Fig. \ref{adaptivebeta}) remains constant for each look direction.

\section{Simulation Results}
\begin{figure}[H]
\centering
        \includegraphics[width=80mm]{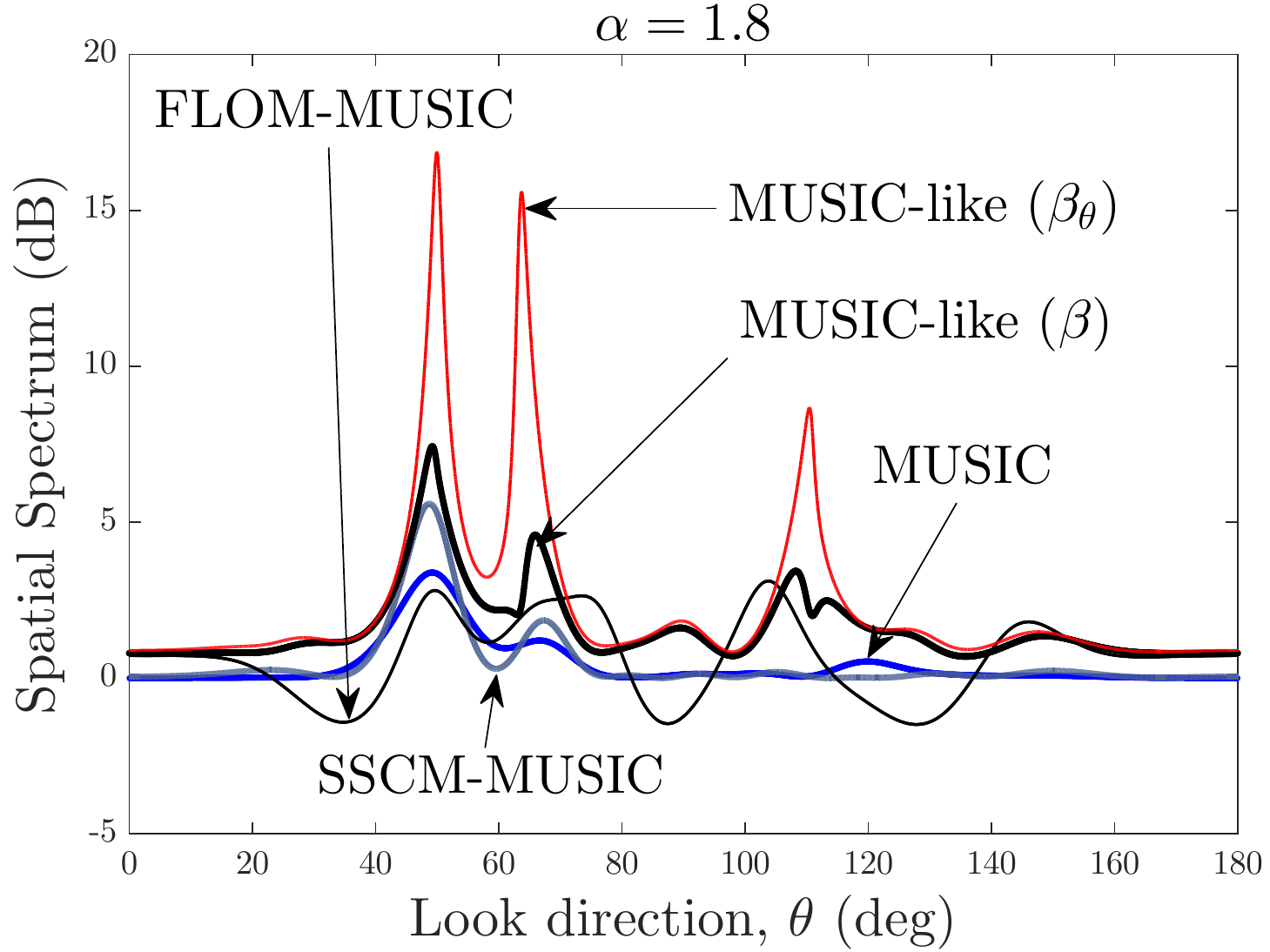}
        \caption{The spatial spectrum of related algorithms with 100 snapshots for $M=10$, $K=3$, and GSNR = -2dB.}
        \label{SpatialSpectrum1}
\end{figure}

Consider a ULA with $\textit{M = 10}$ sensors. Three targets are situated at  $\Theta_K = [50^{\circ}$, $60^{\circ}$, $110^{\circ}]$ where 100 data snapshots were obtained. When $\alpha<2$, a generalized signal-to-noise ratio (GSNR), which is defined as $GSNR(dB) = 10log_{10}(E[|s(t)|^{2}]/\gamma^{\alpha})$, will be used. A complex $S\alpha S$ distributed noise is considered in all simulation studies. When $\alpha=2$, the GSNR reduces to the traditional SNR where $\gamma$ plays the same role as the standard deviation $\sigma$ in a Gaussian process. To obtain the FLOM matrix, $p=1.1$ is set according to the best result reported in \cite{FLOM}. In Fig. \ref{SpatialSpectrum1}, $\alpha$ is set to 1.8 where the GSNR is set to -2dB. It can be seen that the MUSIC-like algorithm with adaptive $\beta_{\theta}$ has the ability to differentiate closely spaced sources better than the MUSIC, FLOM-MUSIC, SSCM-MUSIC, and the original MUSIC-like algorithm with fixed $\beta$. 

\begin{figure}[H]
\centering
        \includegraphics[width=90mm]{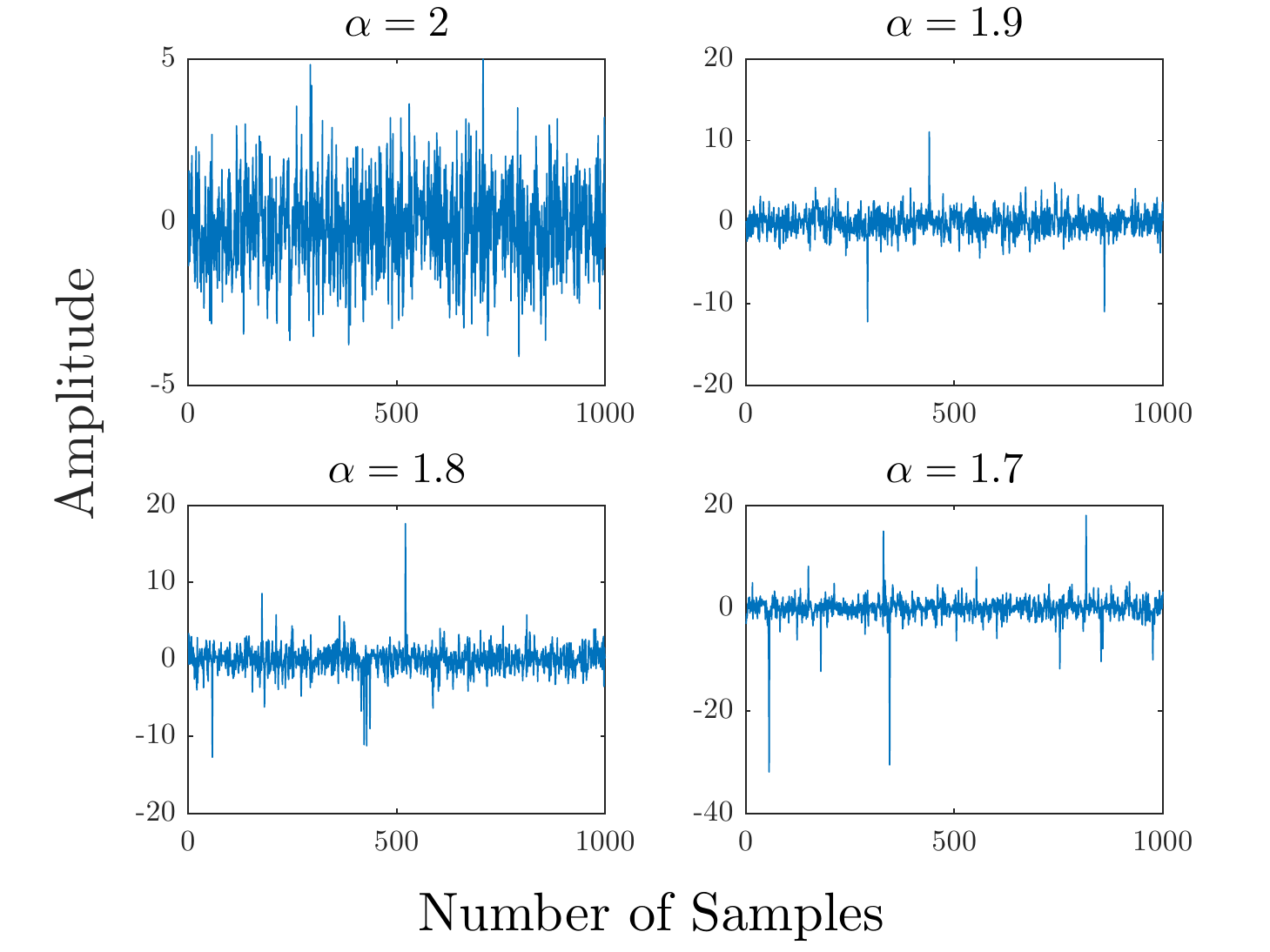}
    \caption{Real part of the complex isotropic $S\alpha S$ distributed noises for $\alpha =$ 2, 1.9, 1.8, and 1.7, respectively.}
    \label{Noises}
\end{figure}

\begin{figure}[H]
		\centering
        \includegraphics[width=65mm]{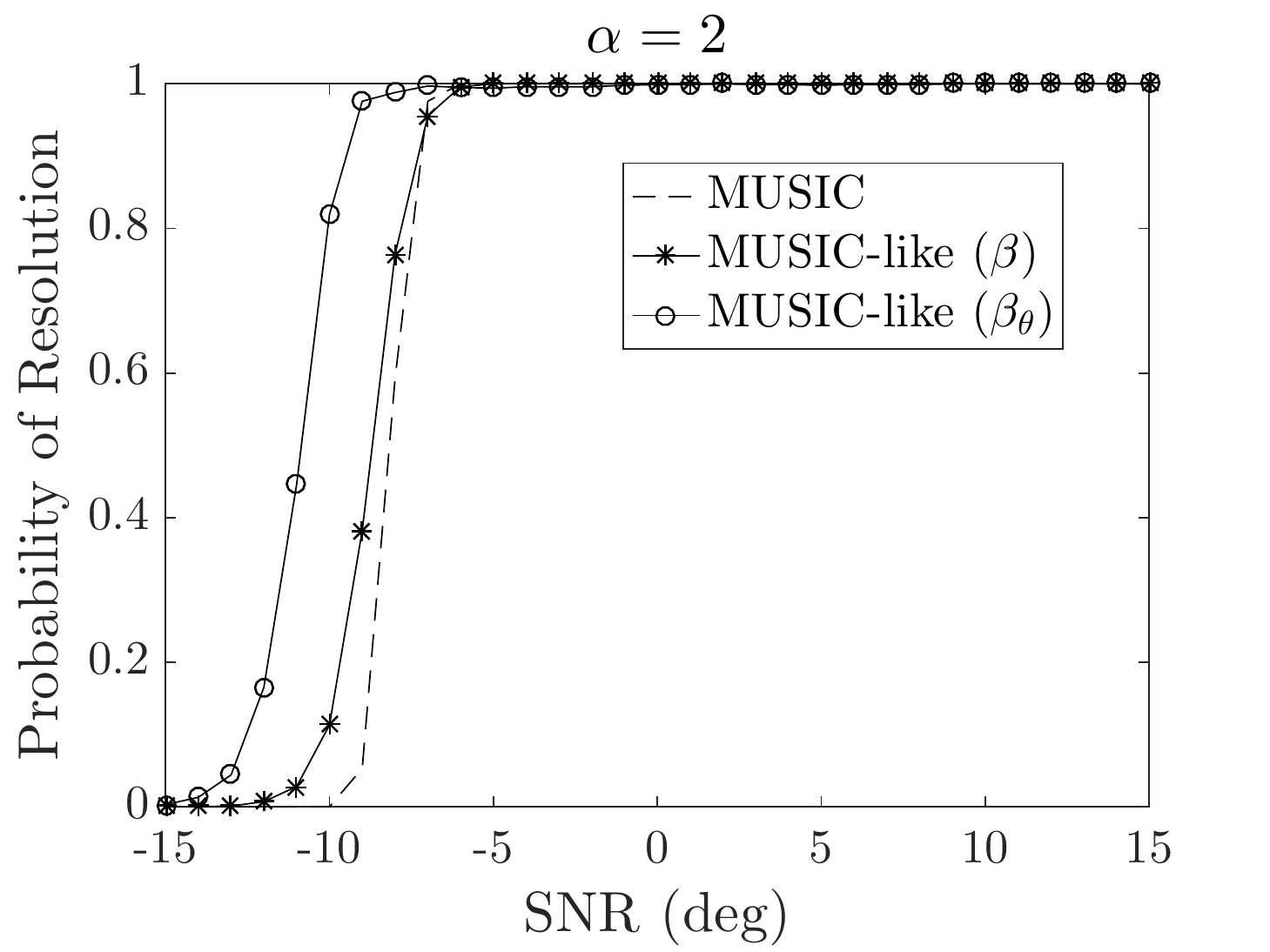}\hspace{-4mm}  
        \includegraphics[width=65mm]{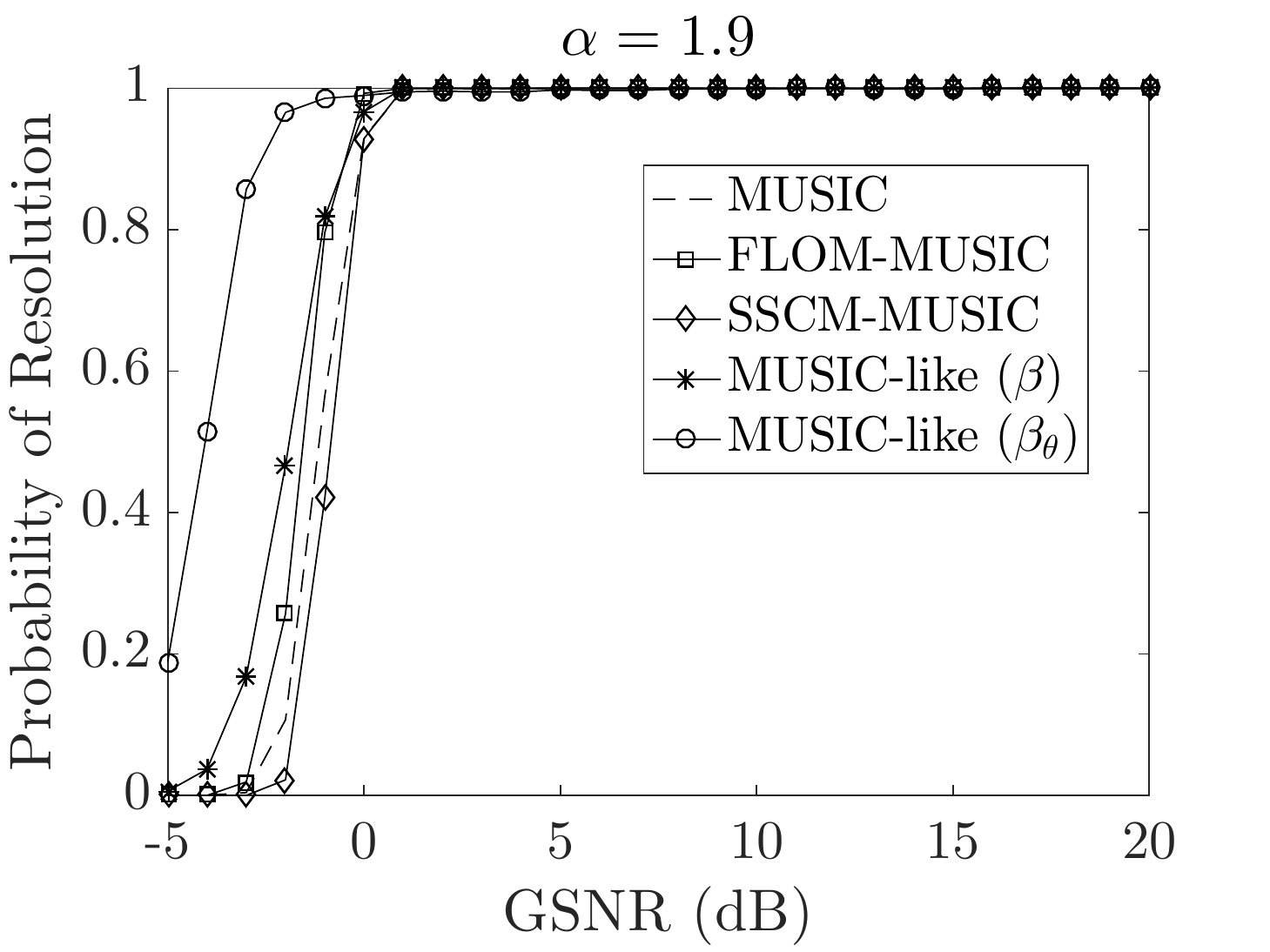}\\    
        \includegraphics[width=65mm]{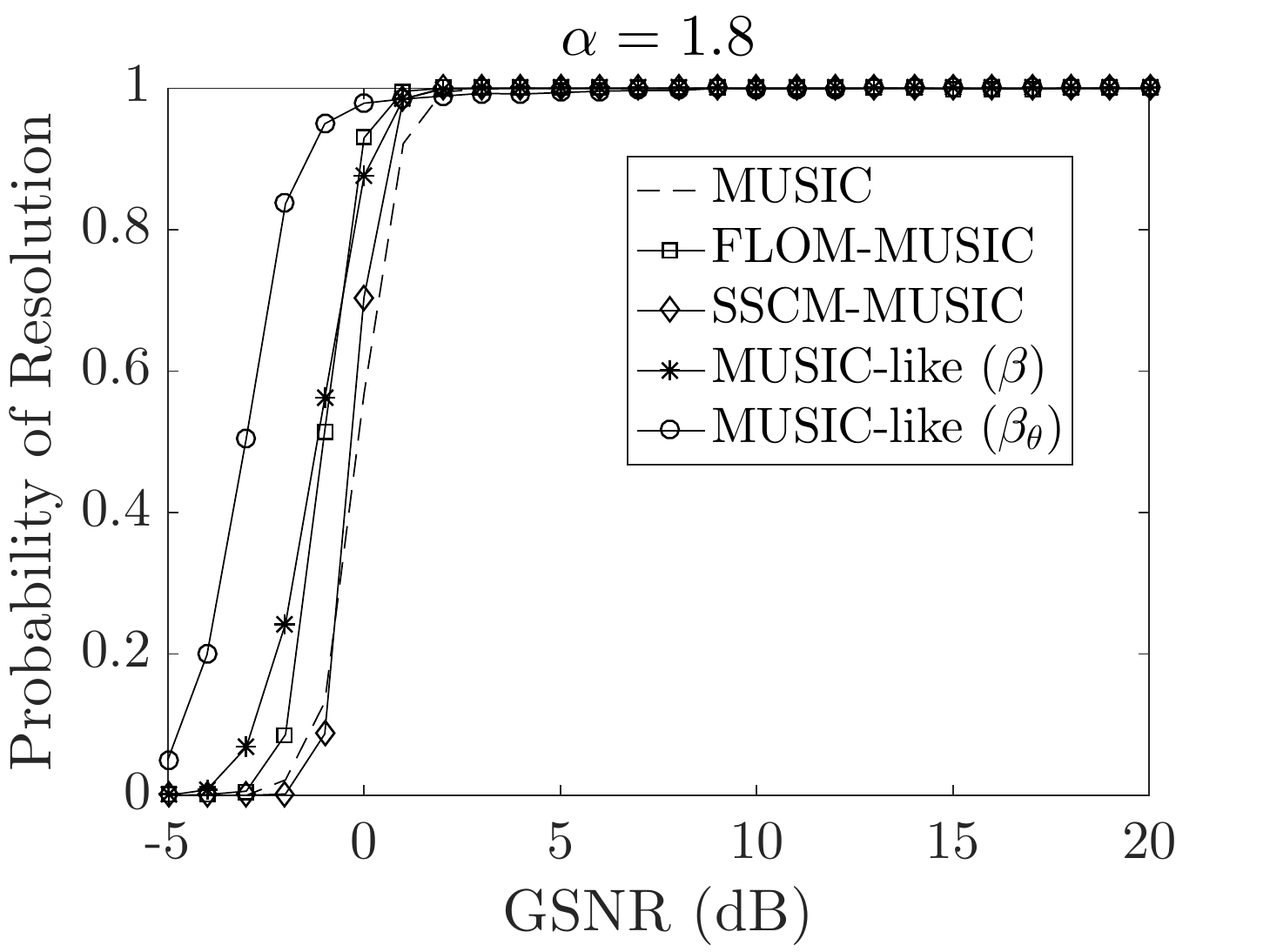}\hspace{-4mm} 
        \includegraphics[width=65mm]{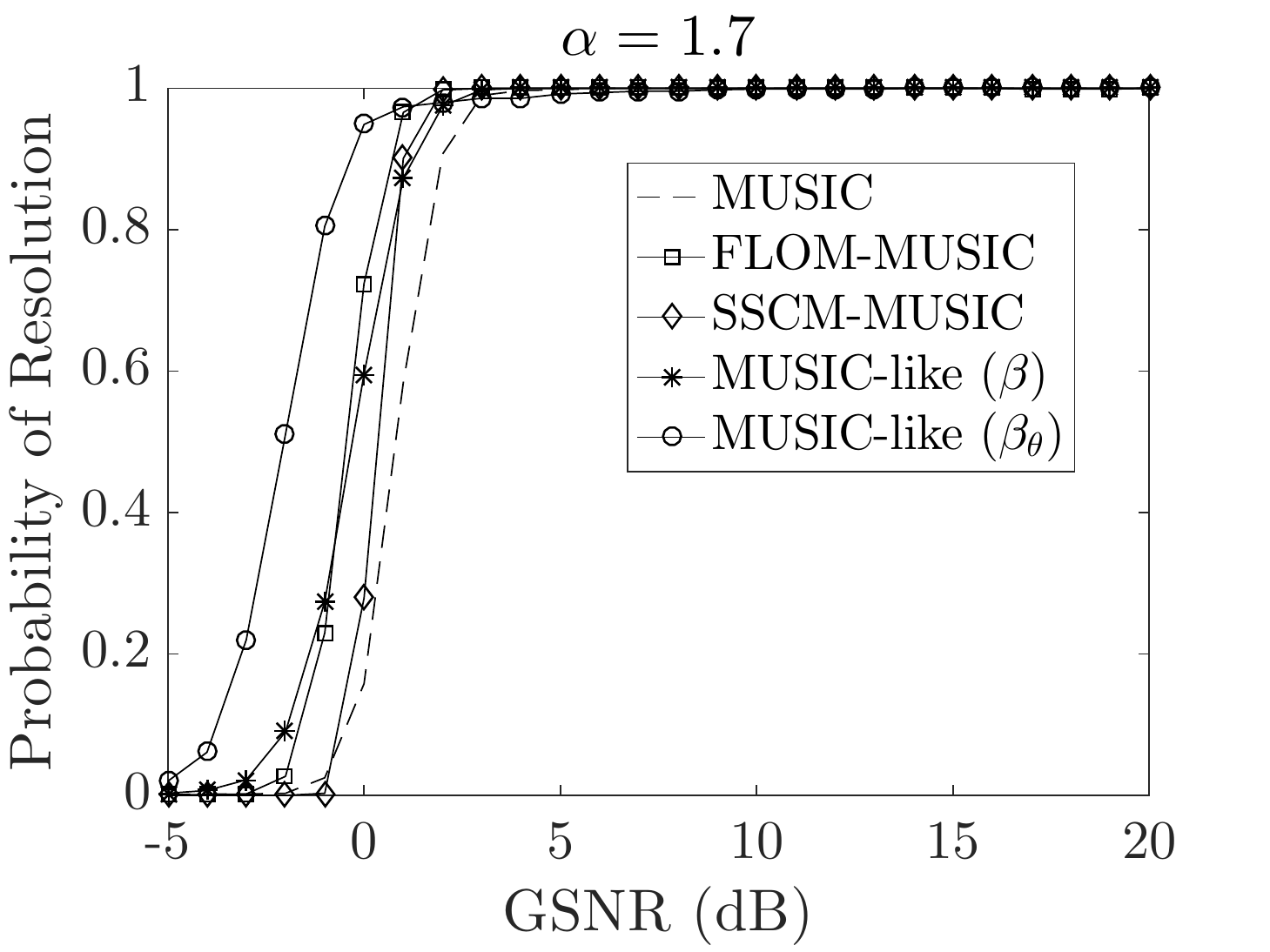}   
    \caption{The probability of resolution plotted against different levels of GSNR with $\alpha=$ 2, 1.9, 1.8, and 1.7, respectively.}\label{Prob}
\end{figure}

\begin{figure}[H]
		\centering
		\includegraphics[width=65mm]{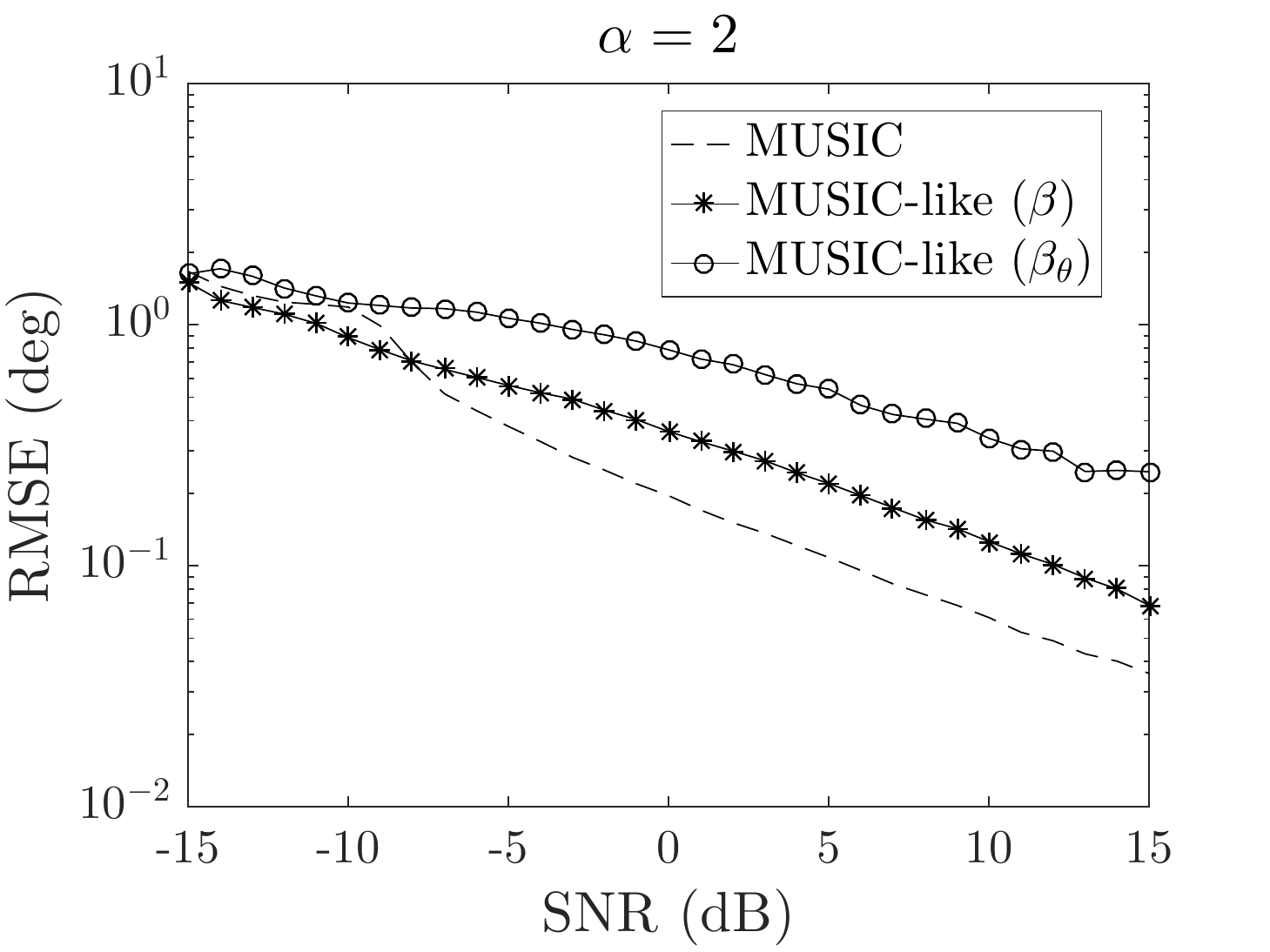}\hspace{-4mm} 
        \includegraphics[width=65mm]{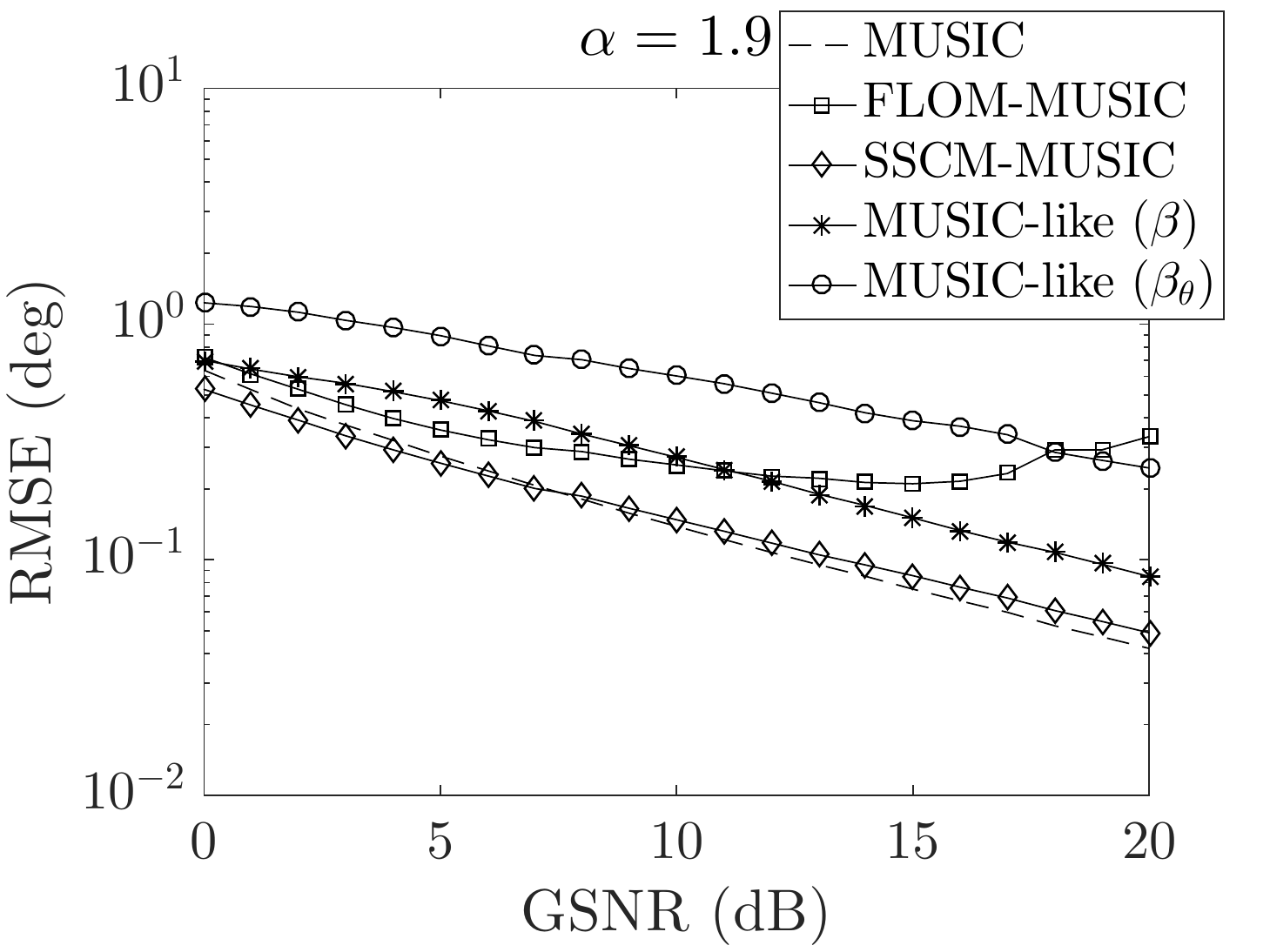}\\
        \includegraphics[width=65mm]{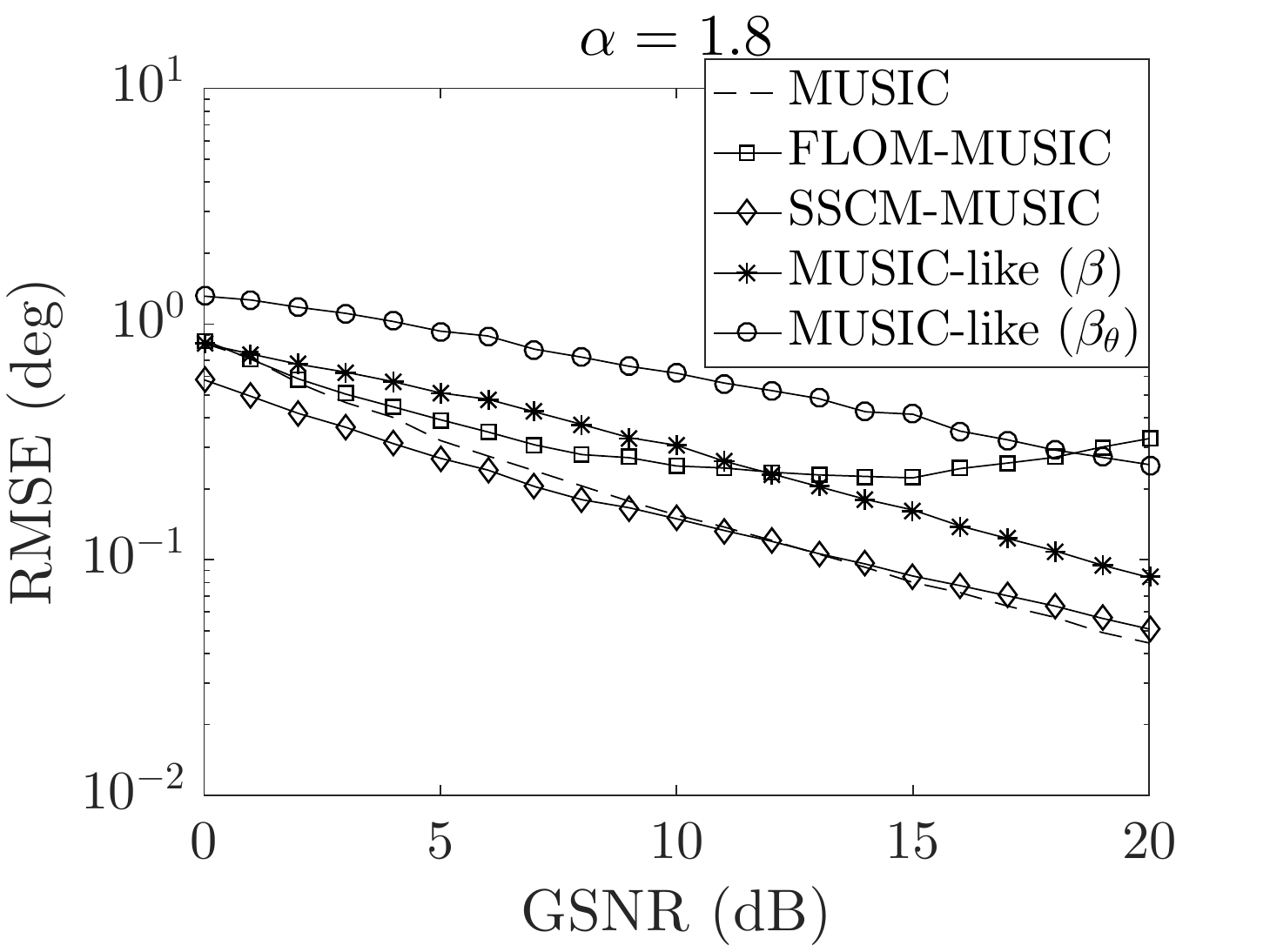}\hspace{-4mm} 
        \includegraphics[width=65mm]{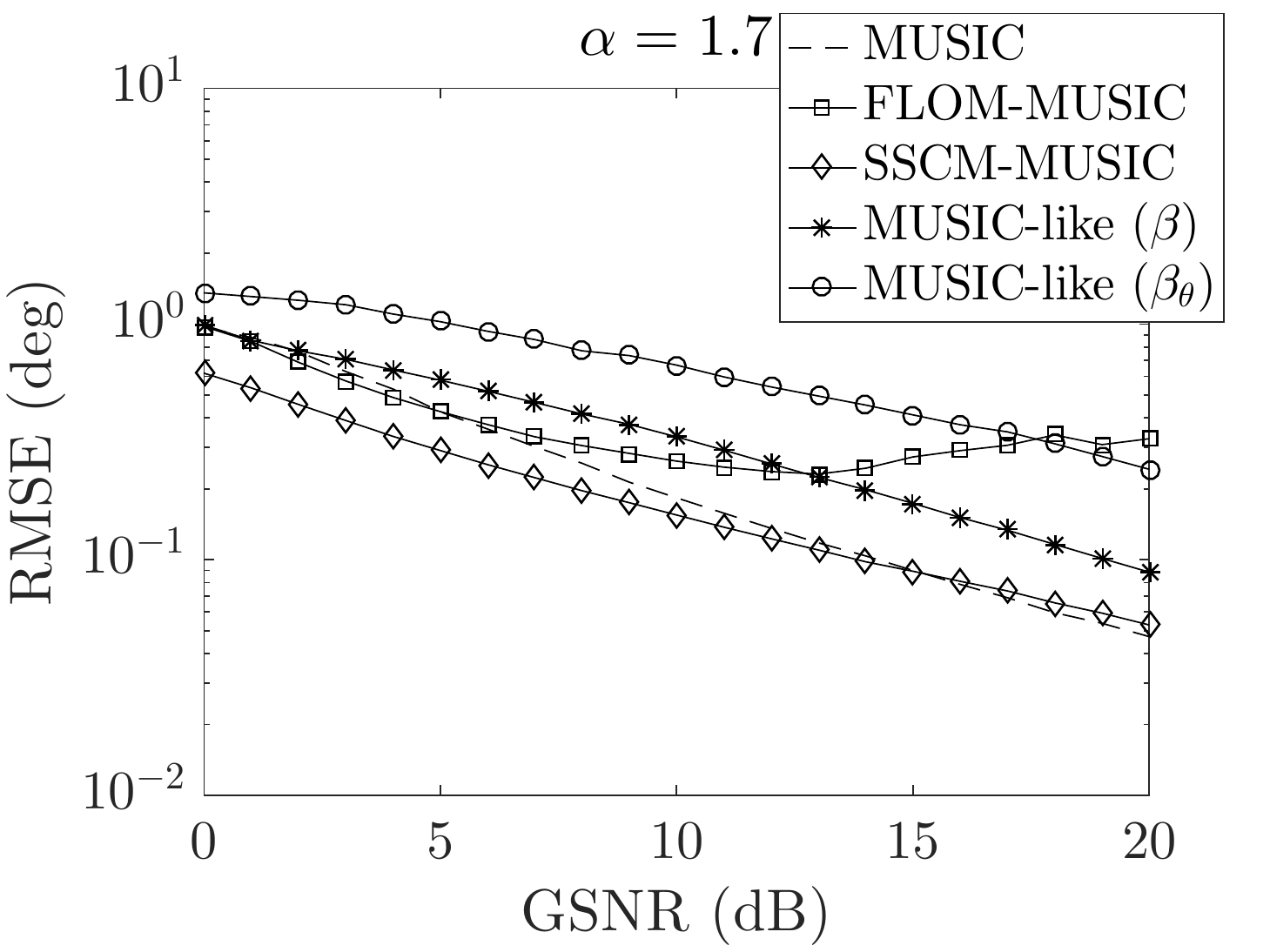}     
    \caption{The average RMSE plotted against different levels of GSNR with $\alpha=$ 2, 1.9, 1.8, and 1.7, respectively.}\label{RMSE}
\end{figure}

Next, performance parameters (probability of resolution and average root-mean-squared error (RMSE)) of related algorithms are studied under the Monte Carlo simulation of 1,000 trials based on different values of $\alpha$. To illustrate the effect of each selected $\alpha$, the real part of complex isotropic $S\alpha S$ distributed noise corresponds to $\alpha =$ 2, 1.9, 1.8, and 1.7 are shown in Fig. \ref{Noises}. To obtain performance parameters, three targets are situated at  $\Theta_K = [50^{\circ}$, $65^{\circ}$, $110^{\circ}]$ where 100 snapshots of data were obtained. The probability of resolution and the average RMSE with different levels of $\alpha$ are shown in Figs. \ref{Prob} and \ref{RMSE}. When $\alpha=2$, the $S\alpha S$ distribution reduces to a Gaussian process, and as $\alpha$ decreases ($\alpha<2$) the likelihood of outlier occurrence increases accordingly. It can be seen in Fig. \ref{Prob} that the proposed method is able to achieve very good performance for the probability of resolution (best targets resolvability) especially in low GSNR conditions. Slight improvement from MUSIC algorithm can be obtained from the SSCM-MUSIC especially when the value of $\alpha$ decreases (e.g. $\alpha = 1.7$). Performance of the original MUSIC-like algorithm with fixed $\beta$ is comparable to the FLOM-MUSIC.


\noindent Although with significant improvement on targets resolvability, we also note a crucial trade-off between such improvement and the estimation RMSE. It can be seen in Fig. \ref{RMSE} that the proposed method tends to have higher bias than the other methods on estimated directions. Such trade-off would be tolerable in certain applications such as the source localization under quasi-static environment, where detection and resolvability are of highest priority and certain biasness is acceptable \cite{DOT,Transcan,EIT}.


\section{Conclusion}
In this letter, geometrical interpretation of the MUSIC-like algorithm was provided which helps to understand the working principle of relaxation parameter $\beta$. Investigations of the original MUSIC-like algorithm and the proposed method under $S\alpha S$ distributed noise were conducted. The proposed method was also compared with the MUSIC, FLOM-MUSIC, and the SSCM-MUSIC algorithms. Computer studies highlight a notable improvement of the proposed method over other methods in terms of targets resolvability. We also note a crucial trade-off between such improvement and the estimation bias which is inherent in the proposed method. The proposed method is therefore suitable for applications where detection and targets resolvability are of highest priority and slight bias is acceptable.

\bibliographystyle{IEEEtran}
\bibliography{References}

\end{document}